\documentstyle[12pt]{article}
\newcommand{\beq}{\begin{equation}}
\newcommand{\eeq}{\end{equation}}

\newcommand{\dd}{D\hspace{-.65em}/}

\def\dop{Dirac operator}

\def\cn{condition}

\begin{document}
\begin{titlepage}
\begin{flushright}
NBI-HE-93-74 \\
December 1993\\
\end{flushright}
\vspace{0.5cm}
\begin{center}
{\large {\bf Topological Classification \\
of Odd-Parity Sphaleron Deformations}}\\
\vspace{1.5cm}
{\bf Minos Axenides}
\footnote{e-mail:axenides@nbivax.nbi.dk}\\
\vspace{0.4cm}
{\em The Niels Bohr Institute\\
University of Copenhagen, 17 Blegdamsvej, 2100 Copenhagen, Denmark}\\
\vspace{0.4cm}
{\bf Andrei Johansen}
\footnote{e-mail:johansen@lnpi.spb.su}\\
\vspace{0.4cm}
{\em The St.Petersburg Nuclear Physics Institute\\
Gatchina, St.Petersburg District, 188350 Russia}\\

\end{center}
\begin{abstract}

We discuss topological aspects
of the electroweak sphaleron and its odd-parity deformations.
We demonstrate that they are uniquely classified in terms of their
odd and even-parity pure gauge field behaviour at spatial infinity.
Fermion level crossing occurs only for odd-parity
configurations which are topologically disconnected from the vacuum.
They contribute to the high temperature unsuppressed
baryon violating thermal transitions.
Deformations with even-parity pure gauge behaviour at spatial
infinity are topologically trivial and
do not mediate baryon number violation in the early Universe.

\end{abstract}
\end{titlepage}
\newpage
\section{Introduction}
\setcounter{equation}{0}

The standard model of electroweak interactions violates baryon number
through the chiral anomaly \cite{hooft}.
The anomaly of the chiral current becomes an anomaly in the baryon and
lepton currents because the electroweak theory is chiral.
At zero temperature $B$-violating transitions are exponentially
suppressed.
At high temperature it has been convincingly argued \cite{cohen}
that the transition rates are dominated by the sphaleron \cite{klink}.
It is a finite energy saddle point solution to the electroweak equations
of motion that has one unstable mode.
This is because it is at the highest point of a continuous set of
configurations that interpolates between the topologically
distinct vacua with Chern-Simons (CS) numbers $n$,
$n+1$ ($n\in {\bf Z}$) and has CS$=n+1/2.$
In thermal equilibrium and at temperature $T\leq E_{sp}$ the probability
of forming a coherent sphaleron configuration in the hot plasma is given
by the Boltzman weight of the classical sphaleron energy ($\Gamma
\propto B \exp \frac{-E_{sp}}{T}$) \cite{peter}.

In the presence of high temperature fluctuations of the gauge and Higgs
fields transitions across the sphaleron saddle point are
accompanied by fermion level crossing.
This formally manifests itself with the presence of normalizable zero energy
solutions to the three dimensional Dirac equation in the sphaleron
background \cite{kunz}.
These transitions are expected to become
unsuppressed at $T\approx E_{sp}$ where perturbative estimates loose
their validity.
Strictly speaking in an incoherent thermal plasma any of
the nearby configurations to the sphaleron with energy
$(E \geq  E_{sp}$) become likely to be generated spontaneously.
This would result from a transition rate
estimate in terms of the Gibbs free energy of all such nearby
configurations at the top of the barrier instead of just the classical
energy of a single sphaleron.
These would include deformed sphaleron
configurations which are not solutions to the electroweak equations of
motion but nevertheless allow for fermion level crossing in their
backround.
We will henceforth generically name such configurations
sphaleron deformations.

Fermion level crossing is therefore not restricted only to sphaleron
configurations \cite{klink} or to deformed sphaleron
solutions of the electroweak equations \cite{yaffe}.
Sphaleron deformations define the general class of sphaleron-like
gauge and Higgs field configurations (non- solutions)
in whose background the Dirac
equation admits normalizable zero energy solutions or equivalently
allows for fermion level crossing.
The existence of such configurations, for example, with CS$\neq 1/2$
and purely induced by the
Yukawa interactions has been recently
demonstrated both analytically \cite{ajn} and numerically \cite{ian}.
In a first order electroweak phase transition possible departure from
thermal equilibrium in the broken phase behind the expanding bubble
walls have also been envisioned \cite{turok}.
This would result from the
liberated latent heat of conversion of the false symmetric vacuum into
the true vacuum of spontaneously broken symmetry.
Sphaleron-like fluctuations are expected to dominate thermal
transition rates away from equilibrium.

It is expected therefore, one way or another,
that in the hot electroweak plasma
sphaleron deformations with equal or higher energy to that of the
sphaleron will also contribute to the rapid baryon violating
transition rates.
Indeed their presence is especially likely to dominate the
symmetric high temperature phase of the electroweak theory $(T\geq
E_{sp}\approx M_w/\alpha)$.
In fact in this regime scaling arguments
indicate unsuppressed baryon violating transition rates $(\Gamma\propto
T^4)$ \cite{dine}.
Equivalently there is an infinitude of paths through
sphaleron deformations that connect two adjacent vacua.
Recent computer simulations of the hot electroweak sphaleron transitions
in the symmetric phase corroborate to this physical picture \cite{ian}.
In such
a nonperturbative phase and in the presence of large thermal fluctuations of
the gauge and Higgs fields sphaleron
deformations dominate fermion level crossing.
For definiteness in what follows we will focus on the ones with CS=1/2.
They lie on the ridge of the saddle point of the sphaleron and have
two unstable modes.
The first brings about their rolling down towards the sphaleron and
the second towards the vacuum.
In the present work we attempt to gain a better understanding of the
general properties of such sphaleron deformations.
We do it by establishing
sufficient conditions in order that a general deformation of the sphaleron
is topologically nontrivial and equivalent homotopically to it.
We do it for the case of the spherical electroweak sphaleron, i.e. in
the limit of zero weak mixing angle.

The paper is organized as follows.
In section 2 we give a topological classification for
odd-parity gauge fields with pure gauge behaviour at spatial infinity.
In section 3 we make explicit constructions of odd-parity sphaleron
deformations by using twisted loops of the electroweak
Nielsen-Olesen vortex.
We use our scheme to classify the latter and identify the ones that are
topologically equivalent to the sphaleron.
We conclude by discussing the possible cosmological role odd-parity
sphaleron deformations might have played in baryogenesis at the
electroweak phase transition in the early Universe.

\section{A Topological Classification}
\setcounter{equation}{0}

We start with the observation that our familiar static
sphaleron configuration has an odd-parity gauge field everywhere in
space.
By imposing the same property on all possible deformations (they may not
be solutions) we find two topologically distinct sectors of
configurations that depend on the (even-odd) parity properties
of their pure gauge behaviour at spatial infinity.

Let us recall some properties of the sphaleron configuration.
Its gauge field reads
\beq
W_k = f(r) \frac{\epsilon_{ijk} x^i \tau^j}{r^2}
= -if(r) \partial_k U_{sph} \; U_{sph}^{-1} ,
\eeq
where $k =1,2,3$ and
\beq
U_{sph} = \frac{ix_i \tau_i}{r}.
\eeq
The function $f(r) \to 1$ at $r\to \infty$
so that this configuration is purely gauge at infinity.

The Chern-Simons functional is defined to be
\beq
CS(W) = \frac{1}{32\pi^2} \int_{D^3} {\rm Tr}(W dW - \frac{2i}{3} W^3)
\eeq
and for the sphaleron equals 1/2.
This fact can be checked almost without calculation due to
the observation that this gauge field configuration is odd under parity.
One can define correctly the Chern-Simons number for the
sphaleron by making a gauge transformation in order to remove the field
at infinity.
Such a transformation is given by an $SU(2)$ group element $U'$
which is smooth everywhere and coincides with $U_{sph}$ at infinity.
Since $\pi_2 (SU(2)) =0$ we know that such a field $U'$ does exist.
The Chern-Simons number for the sphaleron is defined
as the functional ${\rm CS} (W')$, where
\beq
W'_k = U' W_k (U')^{-1} +i \partial_k U' \; (U')^{-1} .
\eeq
We observe that
\beq
{\rm CS} (W') ={\rm CS} (W) + S_{WZW} (U') ,
\eeq
where CS stands for the Chern-Simons functional while
$S_{WZW}$ is the Wess-Zumino-Witten functional defined as follows
\beq
S_{WZW}(U') = \frac{1}{24\pi^2} \int_{D^3} {\rm Tr} (dU'\; U'^{-1})^3 .
\eeq
The latter term actually  depends on only the behaviour of the field $U'$
at the boundary $S^2$ of the disk $D^3$ and is equal to 1/2 for the
particular field $U'$ introduced above.
For example we can take
\beq
U' = \exp \frac{i\pi}{2} \frac{\tau_i x_i}{\sqrt{x^2 + \rho^2}}
\eeq
and check by an explicit calculation that $S_{WZW} (U') =1/2 .$
In turn ${\rm CS} (W) =0$ since the field $W$ is odd under parity, i.e.
$W(-x) = -W(x).$
Thus we conclude that ${\rm CS} (W') =1/2.$
It is now clear that the same value of the Chern-Simons functional
corresponds to all odd-parity configurations with this particular
behaviour at infinity.

Thus we see that the odd parity of the spherical
sphaleron is a very important property.
In fact our short computation indicates that it is necessary
for the Chern-Simons number of the sphaleron to be exactly 1/2.
However it is not sufficient by itself and we have to also make use
of the odd parity of $U'$ at infinity.
It is the purpose of the present work to establish
a connection between the Chern-Simon number of gauge fields
and the parity
property of their pure gauge behaviour (i.e. the $U'$ field itself)
at infinity.
Actually we shall argue below that a restriction to odd-parity
gauge field configurations allows us to introduce a useful
topological classification for all of these fields.

We consider the gauge fields which are purely gauge
at infinity (i.e. on the boundary of a
3-dimensional ball, which is $S^2$)
\beq
A_i = -i(\partial_i U)\; U^{-1},
\eeq
where $U$ belongs to the $SU(2)$ group.
A restriction of this field $U$ to the boundary $S^2$
of the 3-dimensional ball is a map of $S^2$ into $SU(2) .$

The homotopic group $\pi_2 (SU(2))$ is trivial and hence
all such configurations in the 3-dimensional ball are contractible to
unity.
We now restrict ourselves to the space of 3-dimensional
odd-parity gauge fields.
We want to argue that in this space alone
there still exists a relevant non-trivial homotopic classification.
Indeed let us consider an odd-parity configuration
\beq
A_i (-x) = - A_i (x).
\eeq
On the $S^2$ boundary this is a pure gauge so that
\beq
(\partial_i U) (-x) U^{-1} (-x) = -
(\partial_i U) (x) U^{-1} (x) .
\eeq
It is easy to show that the field $U(x)$ can only be either odd or
even under parity.
Indeed, let us consider the following equation
\beq
(\partial_i -iA_i (x) )^2 \phi (x) = 0
\eeq
where $\phi$ is an $SU(2)$ doublet.
Due to the odd parity of the gauge field $A_i$ the parity conjugated
doublet $\phi (-x)$ is also a solution to the above equation.
On the $S^2$ sphere at infinity this solution behaves like
\beq
\phi (x) = U(x) \phi_0,\;\;\;
\phi (-x) = U(-x) \phi_0 = U(x) \phi'_0 ,
\eeq
where $\phi_0$ and $\phi'_0$ are nonvanishing constant doublets.
The matrix $U(-x)^{-1} U(x)$ is a non-degenerate constant
$SU(2)$ matrix since it maps a constant doublet to
another one.
Thus we get
\beq
U(-x) = U(x) V,
\eeq
while $V$ is a constant matrix of $SU(2) .$
This equation is valid for any point $x$ on  $S^2 .$
Hence changing $x \to -x$ we get
\beq
V^2 =1 .
\eeq
A short inspection now shows that the matrix $V$ should belong to
the centre of $SU(2)$. In equations
\beq
V= \pm 1 .
\eeq
Thus we get two different classes of odd-parity gauge fields:
those with odd $U$ and those with even $U .$

This conclusion reflects a non-triviality of the homotopic
group of maps from the projective sphere
${\rm {\bf R}}P^2 = S^2 /{\bf Z}_2$ (where ${\bf Z}_2$ is a group of
parity reflections with respect to some point in 3-dimensional
space) to the group
$SO(3) = SU(2) /{\bf Z}_2$ (where ${\bf Z}_2$ is the centre of $SU(2)$).
In short $\pi ({\bf R} P^2 ,SO(3)) = Z_2 $.
The consideration above shows that
the odd-parity gauge fields split into two topologically
disconnected equivalence classes.
In other words it is not possible to get from one to the other
continuously through odd-parity gauge field configurations.

Actually the classification of these gauge fields is more complicated.
The above homotopic information also implies the following:
if we extend an odd-parity field $U(x)$ from the boundary $S^2$
into the 3-dimensional ball
then it will encounter a singularity at some point.
As we saw above the sphaleron gauge field is (in an appropriate
gauge) odd under parity and has an odd $U$ and
hence it is disconnected from the vacuum configurations (see eq.(2.2)).
In turn configurations with even $U$ fields are continuously connected to the
vacua $A_i^{n}= i(\partial_i U_n)
U_n^{-1}$ where the group element $U_n$ is given by the even-parity
(at infinity) group elements \cite{ryder}
\beq
U_n = \exp (in\pi \frac{\tau_i x_i}{\sqrt{x^2 + \rho^2}}) ,
\eeq
with $\rho$ being a constant parameter and $n$ an integer.
Vacuum configurations are associated with the different
Chern-Simons numbers given by $n.$
Such a classification is a manifestation of the non-triviality of the
homotopic group $\pi_3 (SU(2)) = {\bf Z}.$
The group element $U_n$
is a constant matrix at infinity and hence it corresponds to a
compactification of $D_3 $ into $S^3.$
In turn by taking into
consideration the group elements which are odd under parity at infinity we
compactify $D^3$ into ${\bf R}P^3.$
Thus the relevant homotopy group
is in this case $\pi({\bf R}P^3, SO(3)) = {\bf Z}+{\bf Z_2}.$
A common feature of all odd-parity, even-$U$ configurations is that they
have an integer valued Chern-Simons functional.
Indeed similarly with the case of the sphaleron we can make a nonsingular
gauge rotation so that we remove the gauge field at infinity.
The Wess-Zumino-Witten functional would give us a Chern-Simons
number for the gauge field configuration.
It is easy to see that the value of $S_{WZW}(U)$ is
invariant under even-parity
smooth deformations of the $U$ field at infinity.
Indeed a variation of the Wess-Zumino-Wess functional reads
\beq
\delta S_{WZW} = \frac{1}{8\pi^2} \int_{D^3} d {\rm Tr}
((U^{-1} \delta U)(U^{-1}dU)^2 ).
\eeq
Since the variation of the group element on the surface $S^2$ is
odd under parity
and its value depends only on the values of the fields
at the boundary we immediately conclude that the present variation of the
Wess-Zumino-Wess functional equals zero.
On the other hand let us consider
a product of even-parity (at infinity) group elements $U_1$ and $U_2$ that
correspond to any two such gauge fields.
We have
\beq
S_{WZW} (U_1 U_2) = S_{WZW} (U_1) +S_{WZW} (U_2)
+ \frac{1}{8\pi^2} \int_{D^3} {\rm Tr} d((U^{-1}_1 dU_1 ) (dU_2 \; U_2^{-1})).
\eeq
The third term in the left hand side of this equation equals zero
due to the odd parity of the integrand at infinity.
Hence we see that the Wess-Zumino-Witten functional acts as a homomorphism
from the group of maps $U$ to a discrete
subgroup of the group of real numbers which is obviously
isomorphic to ${\bf Z} .$
As we argued before the even-$U$ (at infinity) group element
is contractible to the vacuum.
In turn as it is well known that the vacuum can have any integer value of the
Chern-Simons number we conclude that even-parity $U$-fields are indeed
classified by ${\bf Z}.$

Thus all odd-parity even-$U$ gauge fields split into
an infinite set of disconnected equivalence classes which are
labeled by integer values of their Chern-Simons numbers.

Let us now consider odd-parity odd-$U$ gauge fields.
A similar argument shows that the value of the Chern-Simons functional
is a topological invariant while the Wess-Zumino-Witten functional
maps the odd-parity $U$ fields to a discrete subgroup of the group of
real numbers according to eq.(2.18).
On the other hand a product of two odd-parity group elements
$U_1$ and $U_2$ is even under parity.
By taking also into account that the sphaleron has
$S_{WZW} (U) =1/2$ we conclude
that the odd-parity odd-$U$ gauge fields have half-integer
values of the Wess-Zumino-Wess functional and hence are classified
by $n+1/2$ ($n\in {\bf Z}$)
while these equivalence classes are themselves topologically
disconnected one from the other for different values of $n.$

Thus we see that the Chern-Simons functional plays the role of
a topological charge: it takes values in ${\bf Z}$
for even-$U$ and in ${\bf Z} + 1/2$ for odd-$U$ fields respectively.

An immediate implication of such a topological index for
the fermionic spectrum of a 3-dimensional \dop \ is the following.
Let us consider a \dop \ $\dd = \gamma_i
(\partial_i -iA_i)$ in an external odd-parity gauge field $A_i .$
Its non-zero eigenvalues are paired up ($\lambda, -\lambda$).
Hence when the external field varies continuously the number of
zero modes of the \dop \ is invariant modulo 2.
For the sphaleron background this topological invariant is equal to one
while for the vacuum its value is zero.
This means that it is not possible to get to the vacuum from
the sphaleron configuration continuously through odd-parity
gauge field configurations.

{}From the above considerations we conclude that in the presence of
an odd-parity external gauge field the number of fermionic zero modes
is 0 mod 2 for even-$U$ and 1 mod 2 for odd-$U$ configurations.

In effect only odd-parity gauge fields with odd-$U$ pure gauge behaviour at
spatial infinity mediate rapid anomalous baryon violating thermal
transitions in the early universe.
Our classification certainly does not exhaust the space of all sphaleron
deformations.
Yet in its restrictiveness and simplicity it provides us with
sufficient conditions so that we can construct explicitly genuine
``sphaleron-like'' configurations.
We now proceed to give concrete examples as an illustration of this point.
To that end we construct vortex-like
deformations of the spherical sphaleron which are simultaneously odd-parity
in their gauge field and odd-$U$ in their pure gauge behaviour at infinity.
This would be sufficient in order that fermion level crossing is manifest.

\section{Nielsen-Olesen Sphaleron Deformations}
\setcounter{equation}{0}

It has recently been speculated that electroweak W and Z vortex solutions
\cite{nambu} might have mediated rapid baryon violation in the electroweak
phase transition \cite{vach}.
These are unstable solutions \cite{peri}
to the classical electroweak equations of motion of the Nielsen-Olesen
type \cite{holger}.
The possibility that such configurations, if stable, might have
contributed substantially to the baryon assymetry of the universe in a
second order electroweak phase transition has also been considered \cite{bob}.
We will demonstrate the usefulness of our
topological classification by constructing configurations that are
topologically equivalent to the sphaleron solution and
configurations that are contractible to the
trivial vacuum.
To that end let us consider a $W$ string loop configuration.
A straight segment of a $W$ string corresponds to the following gauge
field
\beq
W_i = -i v(r) \partial_i U \; U^{-1},
\eeq
where
\beq
U= \left( \begin{array}{cc} \cos \theta & \sin \theta \\
-\sin \theta & \cos \theta \end{array} \right) .
\eeq
Here $\theta$ is a cylindrical angle, while $r$ is a radial
cylindrical coordinate.
The function $v$ is defined as follows
\beq
v(r) = 1 \;\;\;  {\rm at} \;\; r\to \infty ,
\eeq
$$v(r) = v_0 r^2/R^2 \;\;\; {\rm at} \;\; r \to 0 .$$
Let us consider a two-dimensional slice of a segment in the plane
perpendicular to its axis.
It is easy to show that a two-dimensional \dop \ has
no normalizable zero modes.

Let us consider a $W$-loop with a rotational symmetry with respect to
the axis perpendicular to the plane of the loop.
The string loop
configuration is a loop of $W$ string twisted by an azimutal angle
$\phi$ with respect to the centre of the loop.
The gauge field of the W-loop is given by eq.(3.1) while
\beq
U = U_{loop} = \exp(ik\phi \frac{1+ \tau_3}{2})\frac{1}
{\sqrt{(r-x_0)^2 +
z^2}}\left( \begin{array}{cc} r-x_0 & z\\ -z & r-x_0 \end{array} \right)
\eeq
and $v= v((r-x_0)^2 +z^2).$
Here $r, \phi$ and $z$ are cylindrical
coordinates and $x_0$ is a positive constant that denotes the
distance (radius) of the loop from its center $x=y=z=0 ;$
$k$ is an integer and plays a role of winding number.
In the above equation we used the twisting prescription given by \cite{vach}.
We observe however that this gauge field has a $U(1)$ component.
As our topological
classification is formulated for purely $SU(2)$ gauge fields
we modify the $W$-loop so that we generate an $SU(2)$ gauge field.
We do it by considering an alternative (assymmetric)
twisting prescription so that the gauge field is given by eq.(3.1)
while instead of $U$ we substitute
\beq
U'_{loop}= \exp (i\phi \tau_3 /2)\frac{1}{\sqrt{(r- x_0)^2 +z^2}}
\left( \begin{array}{cc} r-x_0 & z \\ -z & r-x_0 \end{array} \right)
\exp(-i\phi \tau_3 /2).
\eeq
Such a configuration is single valued, odd under parity, even-$U$ and hence
it is topologically trivial for any profile function $v .$
It is therefore disconnected from the sphaleron configuration and has
an integer Chern-Simons number.
In the presence of this gauge field there is an even number of
normalizable fermionic zero modes.

However there exists a different symmetric twisting that gives an
odd-parity odd-$U$ gauge field configuration
\beq
U'_{loop}= \exp (i\phi \tau_3 /2)\frac{1}{\sqrt{(r- x_0)^2 +z^2}}
\left( \begin{array}{cc} r-x_0 & z \\ -z & r-x_0 \end{array} \right)
\exp(i\phi \tau_3 /2).
\eeq
This confuguration is connected to the sphaleron by the topological arguments
given above and has ${\rm CS}=1/2 .$
The number of fermionic zero modes is odd in this case.

It is clear that it is easy to generalize these two constructions
to a twisting with any odd winding numbers
\beq
U'_{loop}= \exp (i(2n+1)\phi \tau_3 /2)\frac{1}{\sqrt{(r- x_0)^2 +z^2}}
\left( \begin{array}{cc} r-x_0 & z \\ -z & r-x_0 \end{array} \right)
\exp(i(2m+1)\phi \tau_3 /2),
\eeq
where $n$ and $m$ are integers.
For topologically trivial configurations
(CS is integer) $n+m$ must be odd while
for topologically non-trivial ones (CS is half integer) $n+m$ must be even.

Physically the assymmetrically twisted loop can be interpreted as
a bound state of two sphalerons while the symmetrically twisted
one is to be associated to a single deformed sphaleron.
This will be clear if we take the limit of a collapsed loop $(z_0 \to 0)$ for
both cases.
For the symmetrically twisted loop we get
\beq
U'_{loop}(x,y,z) \to
\exp (i\phi \tau_3 /2)\frac{1}{\sqrt{r^2 +z^2}}
\left( \begin{array}{cc} r & z \\ -z & r \end{array} \right)
\exp(i\phi \tau_3 /2) =
\eeq
$$\frac{1}{\sqrt{r^2 +z^2}}
\left( \begin{array}{cc} x+iy & z \\ -z & x-iy \end{array} \right) =
-i\tau_2 U_{sph}(x,y,z) i\tau_3 .$$
Thus the collapsed loop coinsides (up to a profile function) with the
sphaleron configuration.

In the case of a trivially (assymetrically) twisted loop we get
\beq
U'_{loop} \to
\exp (i\phi \tau_3 /2)\frac{1}{\sqrt{r^2 +z^2}}
\left( \begin{array}{cc} r & z \\ -z & r \end{array} \right)
\exp(-i\phi \tau_3 /2) =
\eeq
$$\frac{1}{\sqrt{r^2 +z^2}}
\left( \begin{array}{cc} r & ze^{i\phi} \\ -ze^{-i\phi} & r
\end{array} \right) .$$
We may now introduce a continuous parameter $\alpha.$
By rescaling  $z \to \alpha z ,$
and by continuously taking $\alpha \to 0$ we see that $U'_{loop}$
can be continuously deformed to $U=1$ by preserving its parity properties.
This is an illustration of the topological triviality of
assymmetrically twisted loops.

We now deform the assymmetrically twisted loop into a configuration which
geometrically looks like two loops of equal radii with their
centers on the z-axis at $z_0$ and $-z_0$ respectively.
We assign symmetrical twistings with opposite handedness to both of them.
The topological equivalence of each one of them to the sphaleron implies
an interpretation for the assymetrically twisted topologically trivial
W-loop as a superposition (bound state) of two odd-$U$ sphaleron deformations.
This type of deformation can be understood as follows.
The topologically trivial $W$-loop corresponds to $U_1 = TU_0 T^+ ,$
where $U_0$ stands for the untwisted $U$ matrix in eq.(3.4) at $k=0,$
and $T$ is a twisting exponential $\exp (i\phi \tau_3 /2) .$
The matrices $U_+ =T U_0 T$ and $U_- =T^+ U_0 T^+$ correspond to
topologically non-trivial symmetrically twisted loops with opposite
handedness.
We now consider the trivially twisted loop with $TU_0 T^+ .$
This matrix can  split into a product
\beq
TU_0 T^+ = TU_0^{1/2} T T^+ U_0^{1/2} T^+ .
\eeq
The angle $\theta$ which is an angle going around the string
from $0$ to $2\pi$ splits therefore into two angles $\theta_1$ and
$\theta_2$ which take values in $(0,\pi)$ and $(\pi ,2\pi)$
respectively.
These two angles correspond to the matrices $U_0^{1/2}$ above.
While the $U_0^{1/2}$ is not single valued a product of them can very well be.
By changing the coordinate dependence of $U_0^{1/2}$ and that of
the profile function one can split one loop of radius $r=x_0$ with a
center at $z_0=0$ into two
loops of equal radii and centers separated by $2z_0$ as follows (see Fig.2)
\beq
U'_{loop} \to \exp (i\phi \tau_3 /2)
\left[ \frac{1}{\sqrt{(r- x_0)^2 +(z-z_0)^2}}
\left( \begin{array}{cc} r-x_0 & z-z_0 \\ -z+z_0 & r-x_0 \end{array}
\right) \right]^{1/2}
\exp(i\phi \tau_3 /2) \times
\eeq
$$\exp (-i\phi \tau_3 /2)
\left[ \frac{1}{\sqrt{(r- x_0)^2 +(z+z_0)^2}}
\left( \begin{array}{cc} r-x_0 & z+z_0 \\ -z-z_0 & r-x_0 \end{array} \right)
\right]^{1/2} \exp(-i\phi \tau_3 /2) .$$
The line of zeros of the profile function consequently splits into two
lines with $z=z_0$ and $z=-z_0$ respectively.
In the limit $z_0 \to \infty$ the angles $\theta_{1,2}$
take effectively values in $(0,2\pi)$ near the cores of the separate loops.
Thus these loops can be viewed as distinct single valued
gauge configurations in the limit $z_0 \to \infty .$

It is clear that under splitting in the $z$ direction
the deformation of the assymmetrically twisted loop is not odd under parity.
Indeed under parity conjugation we have to interchange the position
of the two nontrivial sphaleron deformations.
However there is a way to implement an odd-parity preserving
deformation.
We do it by splitting the trivially twisted loop into
two topologically non-trivial and concentric
ones $(z_0=0)$ but with different radii $r$ (see Fig.3).
In this case we have
\beq
U'_{loop} \to \exp (i\phi \tau_3 /2)
\left[ \frac{1}{\sqrt{(r- x_1)^2 + z^2}}
\left( \begin{array}{cc} r-x_1 & z \\ -z & r-x_1 \end{array}
\right) \right]^{1/2}
\exp(i\phi \tau_3 /2) \times
\eeq
$$\exp (-i\phi \tau_3 /2)
\left[ \frac{1}{\sqrt{(r- x_2)^2 +z^2}}
\left( \begin{array}{cc} r-x_2 & z \\ -z & r-x_2 \end{array} \right)
\right]^{1/2} \exp(-i\phi \tau_3 /2) ,$$
where $x_{1,2}$ are the radii of central lines of strings
in the resulting loops.
It is important that we maintain the odd parity of the whole gauge field
under such a deformation.
This is necessary so that we keep the gauge field
in the same topological class as before.

Let us now consider possible deformations of a nontrivial symmetrically
twisted loop (Fig.1b) similar to those of trivial assymmetrically twisted loop.
It is obvious that a deformation into two assymetrically twisted loops
(Fig.1a) is impossible.
It can be easily seen however that an odd-parity
superposition of one topologically trivial W-loop with a nontrivial one is
allowed.
In a similar way a symmetrically twisted loop can
split into an odd-parity superposition of an assymmetrically
twisted (i.e. trivial) loop and a symmetrically twisted one.
We can now make use of the two elementary deformations
presented in order to build non-minimal configurations.
By taking into account
that $T^2 = TU_0^{1/2}T T^+ U_0^{-1/2}T$
it is straightforward to show that
non-minimal twisted loops with $m,$ $n \neq 0$ can be interpreted as
multiple sphaleron-antisphaleron bound states.
Topologically non-trivial twisted loops are
a superposition of an odd number
of elementary non-trivial ones while the trivial loops correspond
to even numbers of deformed sphalerons and antisphalerons.
Notice also that for a general non-minimal twisting
it is possible to get an odd-parity splitting of loops
in the $z$ direction too.

The case of a different twisting of the loop associated with
a shift $m ,$ $n \to m+1/2 ,$ $n+1/2$ does not correspond to
any odd-parity gauge field configuration and therefore it does
not fit into our classification.
We postpone a detailed analysis of a generalization of our
classification for future work.
Here we only want to indicate that this type of configurations
does not have a simple interpretation in terms of deformed
(anti)sphalerons.

\section{Conclusions}
\setcounter{equation}{0}

We have argued that any odd-parity
sphaleron deformation that
contributes to rapid baryon violating transition rates in the early Universe
must be topologically connected to the electroweak sphaleron.
We demonstrated that a sufficient \cn \ for that is that its pure gauge
behaviour at infinity is given by an odd-parity $U$-field.
Deformations which are not odd under parity such as deformed sphaleron
solutions with CS$\neq 1/2$ or the ones induced by Yukawa interactions
\cite{ajn} are beyond the reach of our classification.

In closing we discuss the possible role sphaleron deformations might
have played in the baryogenesis at the electroweak scale.
In the context
of a first order electroweak phase transition the expanding bubble wall
of the broken symmetry phase drives the baryon and CP violating
processes out of thermal equilibrium.
Consequently they are the region
where the biasing of the baryon number effectively takes place \cite{cohen}.
In their absence, such as in a second order
transition, it has been argued that a similar biasing effect can arise
by the moving edges of the electroweak vortex solutions.
If stable they are
expected to have been produced via the Kibble mechanism \cite{bob}.
Our present work suggests that sphaleron deformations could also play a
similar role.
Electroweak vortex solutions constitute a set of measure
zero in the large class of such
configurations that could contribute to baryogenesis.
We would expect, for example, that the evolution of large networks of
odd-parity odd-$U$ $W$-loops and their eventual shrinking and
contraction to render
their phase boundaries with odd-parity even-$U$ configurations, such as the
vacuum, effective regions of biasing for the rapidly produced baryons.
The precise baryon assymetry produced in such a highly nonperturbative
scenario is certainly a challenge to compute and a highly nontrivial
dynamical problem.

\section{Acknowledgments}
\setcounter{equation}{0}

We thank J.Ambjorn, J.Distler, V.Fock, H.B.Nielsen, P.Olesen and
O.Tornkvist
for many useful discussions and
encouragement.
This research is partially supported by a NATO fund.

\newpage
\subsection*{Figure Captions}
\vskip 6pt

$\;\;\;\;$
Fig.1. Examples of topologically trivial (Fig.1a) and nontrivial (Fig.1b)
sphaleron deformations. They are given by assymetrical-symmetrical
twists of the W-loop, hereby depicted by lines with
antiparallel-parallel orientation respectively.

Fig.2. Odd-parity violating split (deformation) of a
topologically trivial (even-$U$)
$W$-loop.

Fig.3. Odd-parity preserving split (deformation) of a
topologically trivial (even-$U$)
$W$-loop.


\begin{thebibliography}{11}

\bibitem{hooft} G.'t Hooft, Phys. Rev. Lett. {\bf 37} (1976) 8; Phys.
Rev. {\bf D14} (1976) 3432.
\bibitem{cohen}V.A.Kuzmin, V.A.Rubakov and M.E.Shaposhnikov,
Phys. Lett. {\bf 155B} (1985) 36;\\
A.G.Cohen, D.B.Kaplan and A.E.Nelson, Ann. Rev. Nucl. Part.
Sci. {\bf 43} (1993).
\bibitem{klink}N.Manton, Phys. Rev. {\bf D28} (1983) 2019;\\
F.Klinkhamer and N.Manton, Phys. Rev. {\bf D30} (1984) 2212.
\bibitem{peter}P.Arnold and L.McLerran, Phys. Rev. {\bf D36} (1987) 581;
ibid. {\bf D37} (1988) 1020; \\
D.Yu.Grigoriev, V.A.Rubakov and M.E.Shaposnikov,
Phys. Lett. {\bf 216B} (1989) 172.
\bibitem{kunz}J.Boguta and J.Kunz, Phys. Lett. {\bf 154B} (1985) 407;\\
A.Ringwald, Phys. Lett. {\bf 213B} (1988) 61.
\bibitem{yaffe}L.G.Yaffe, Phys. Rev. {\bf D40} (1989) 3463.
F.R.Klinkhamer, Phys.Lett. 236B (1990) 187.
\bibitem{ajn}M.Axenides, A.Johansen and H.B.Nielsen,
``On Sphaleron Deformations Induced by Yukawa Interactions'',
Preprint NBI-HE-93-75.
\bibitem{ian}J.Ambj\o rn et al., ``Level Crossing for Hot Sphalerons'',
Preprint NBI-HE-93-62.
\bibitem{turok}Turok N, Imperial preprint TP-91-92-33(1992), in Per
spectives on Higgs physics, p.300;\\
Grigoriev D, Turok N.,Shaposhnikov M,
Phys. Lett. {\bf 275B} (1992) 395.
\bibitem{dine}M. Dine et al., Nucl. Phys. {\bf B342} (1990) 381;\\
S. Yu. Khlebnikov and M. E. Shaposhnikov, Nucl. Phys. {\bf B308} (1988) 885.
\bibitem{ryder}L.H.Ryder. Quantum Field Theory. Cambridge University
Press, 1985.
\bibitem{nambu}Y.Nambu, Nucl. Phys. {\bf B130} (1977) 505;\\
T.Vachaspati, Phys. Rev. Lett. {\bf 68} (1992) 1977.
\bibitem{vach}M.Barriola, T.Vachaspati and M.Bucher, in "Embedded
Defects", TUTP-93-7.
\bibitem{peri}M.James, L.Perivolaropoulos and T.Vachaspati,
Phys. Rev. {\bf D46} (1992) R5232.
\bibitem{holger}H.B.Nielsen and P.Olesen, Nucl. Phys. {\bf B61} (1973) 45.
\bibitem{bob}R.H.Brandenberger and A.C.Davis,
Phys. Lett. {\bf 308B} (1993) 79.

\end{thebibliography}
\end{document}